\documentstyle[epsfig]{aipproc}

\newcommand{\beq}{\begin{eqnarray}}
\newcommand{\eeq}{\end{eqnarray}}

\newcommand{\la}{\langle}
\newcommand{\ra}{\rangle}
\newcommand{\qc}{\la \bar{q}q \ra}
\newcommand{\uc}{\la \bar{u}u \ra}
\newcommand{\dc}{\la \bar{d}d \ra}
\newcommand{\s}{\la \bar{s}s \ra}

\newcommand{\dmdmu}{{\partial m \over \partial \mu}}

\newcommand{\dmkdmu}{{\partial m_K \over \partial \mu}}
\newcommand{\dmkdmus}{{\partial m_K \over \partial \mu_S}}
\newcommand{\dmkdmuv}{{\partial m_K \over \partial \mu_V}}
\newcommand{\dmpidmu}{{\partial m_\pi \over \partial \mu}}
\newcommand{\dmpidmus}{{\partial m_\pi \over \partial \mu_S}}
\newcommand{\dmpidmuv}{{\partial m_\pi \over \partial \mu_V}}

\newcommand{\dqcdmu}{{\partial \qc \over \partial \mu}}
\newcommand{\ducdmu}{{\partial \uc \over \partial \mu}}
\newcommand{\ddcdmu}{{\partial \dc \over \partial \mu}}
\newcommand{\dscdmu}{{\partial \s \over \partial \mu}}
\newcommand{\dqcdmus}{{\partial \qc \over \partial \mu_S}}
\newcommand{\ducdmus}{{\partial \uc \over \partial \mu_S}}

\newcommand{\dscdmus}{{\partial \s \over \partial \mu_S}}
\newcommand{\dqcdmuv}{{\partial \qc \over \partial \mu_V}}

\newcommand{\ddmkdmu}{{\partial^2 m_K \over \partial \mu^2}}

\newcommand{\ddmpidmu}{{\partial^2 m_\pi \over \partial \mu^2}}

\begin{document}
%
\title{$\dmdmu$ in the Nambu--Jona-Lasinio model}
\author{O.~Miyamura and S.~Choe}
\address{Dept. of Physics, Hiroshima University,
Higashi-Hiroshima 739-8526, Japan}
\maketitle
\begin{abstract}

Using the Nambu--Jona-Lasinio (NJL) model we study responses of
the pion and kaon masses to changes in the chemical potential,
$\dmdmu$, at zero and finite chemical potential.
 We find that the behavior of $\dmdmu$ for
 the pion is quite different
 from that for the kaon. Our results can give a clue for future
studies of $\dmdmu$ on the lattice.

\end{abstract}
%
\section*{1. Introduction}

There are several methods to understand the behavior of matter
under extreme conditions of temperature and/or density. One of
them is the lattice QCD. While the structure of QCD at high
temperature has been investigated in detail, little is known about
matter at high baryon density due to the well-known
``complex-action" problem \cite{lattice}. One of possible ways on
the lattice is to simulate the response of a hadron mass to
changes in the chemical potential, $\dmdmu$, at zero chemical
potential ($\mu$ = 0) \cite{taro,miyamura}.

One the other hand, one can use effective models of QCD : e.g.,
the NJL model \cite{njl}. This model has been widely used for
describing the phase transition as well as hadron properties in
hot and/or dense matter \cite{hk94}.

In this work we present the NJL model calculations of $\dmdmu$ for
the pion and kaon. The primary goal of our study is to get the
same quantities which are simulated on the lattice. Of course, the
direct comparison between the lattice data and the NJL model
calculations is not possible because $\dmdmu$ on the lattice is
for the screening mass while for the pole mass in the NJL model.
Nevertheless, we can learn some ideas for future studies of
$\dmdmu$ on the lattice from this effective model calculations.

Following the notation of the lattice simulations we consider two
kinds of the chemical potential. One is the isoscalar $\mu_S$ =
$\mu_u$ + $\mu_s$ for the kaon (or $\mu_u$ + $\mu_d$ for the
pion). The other is the isovector $\mu_V$ = $\mu_s$ -- $\mu_u$ (or
$\mu_d$ -- $\mu_u$). In contrast to the lattice simulations we get
$\dmdmu$ at zero and finite chemical potential within the NJL
model. Then, our study can give information about the role of the
light quark chemical potential and/or the strange quark chemical
potential in hot and/or dense matter.

The paper is organized as follows. In Sec. 2 we introduce some
basic formulas to get $\dmkdmu$ for the kaon in the NJL model, and
show results at zero and finite chemical potential. We present
$\dmpidmu$ for the pion in Sec. 3. In Sec. 4 we summarize our
results and discuss some uncertainties in our calculations.

\section*{2. $\dmkdmu$ in the NJL model}
\label{sec2}

We use the generalized SU(3) NJL model with the anomaly term
\cite{hk94}:
\beq
 {\em L} = \bar{q}(i\gamma \cdot \partial - m)q
 + {1 \over 2} g_S \sum_{a=0}^8 \left[ (\bar{q}\lambda_aq)^2 +
\bar{q}(i\lambda_a\gamma_5q)^2 \right]
 + g_D \left[ {\rm det} ~\bar{q}_i (1 - \gamma_5)q_j + h.c. \right] \ ,
 \label{lag}
\eeq
where $\lambda_a$ are the Gell-Mann matrices and $m$ is a mass
matrix for current quarks, $m$ = diag($m_u$, $m_d$, $m_s$). We
take the following parameters in \cite{hk94}.
\beq \Lambda = 631.4 ~{\rm MeV}, ~g_S\Lambda^2 = 3.67,
~g_D\Lambda^5 = -9.29
\nonumber \\
m_u = m_d = 5.5 ~{\rm MeV}, ~m_s = 135.7 ~{\rm MeV} \ ,
 \eeq
where $\Lambda$ is the momentum cut-off. The third term in
Eq.(\ref{lag}) is a reflection of the axial anomaly, and causes a
mixing in flavors. For example, the constituent quark masses are
given as follows.
\beq
M_u &=& m_u - 2 g_S \alpha - 2 g_D \beta\gamma \ , \nonumber \\
M_d &=& m_d - 2 g_S \beta - 2 g_D \alpha\gamma \ , \nonumber \\
M_s &=& m_s - 2 g_S \gamma - 2 g_D \alpha\beta \ , \label{mass}
\eeq
where $\alpha \equiv \uc$, $\beta \equiv \dc$, and $\gamma \equiv
\s$. It means that a change of $\uc$ results in a change of $\s$,
and vice versa. Then, we can expect a change in the properties of
the observables related with the strange quarks even in the
nuclear matter.

In this work we concentrate mostly on the Case II in \cite{hk94},
where only $g_D$ has a $T$-dependence
\beq
g_D(T) = g_D(T=0) ~{\rm exp} [-(T/T_0)^2]
\label{gd}
\eeq
while other coupling constants and the cut-off are independent of
$T$ and chemical potential (or density). Here, we set $T_0$ = 0.1
GeV taking into account the restoration of $U_A(1)$ symmetry as in
\cite{hk94}. It might be realistic to make the coupling constants
and/or the cut-off dependent on temperature and chemical
potential. However, at present, there is no such an estimate
including all variations in the cut-off and the coupling constants
except for a few estimates of the strength of the anomaly term
$g_D$ \cite{gdt}.

In the mean-field approximation the above Lagrangian leads to the
following gap equation \cite{hk94}.
\beq
 \la \bar{q}_iq_i \ra  =
 2 N_c \sum_p \left( {-M_i \over E_{ip}} f(E_{ip})\right) \ ,
 \label{gap}
\eeq
where $\la \cdot \ra$ means the statistical average and the index
$i$ denotes the $u$, $d$, and $s$ quarks. $N_c$ is the number of
colors and $M_i$ is the constituent quark mass, and $E_{ip} =
\sqrt{M_i^2 + p^2}$. $f(E_{ip}) = 1 - n_{ip} - \bar{n}_{ip}$,
where $n_{ip}$ and $\bar{n}_{ip}$ are the distribution functions
of the $i$th quark and antiquark, respectively.
\beq n_{ip} = {1  \over 1 + {\rm exp}~((E_{ip} - \mu_i)/T)} \ ,
 \nonumber \\
 \bar{n}_{ip} = {1 \over 1 + {\rm exp}~((E_{ip} + \mu_i)/T)} \ .
\eeq

The right-hand side of Eq.(\ref{gap}) is a function of $F ~(\uc,
\dc, \s, \mu_i, T)$. Then, we obtain responses of the quark
condensates $\ducdmu$, $\ddcdmu$, and $\dscdmu$ by differentiating
both sides with respect to $\mu$ at a fixed $T$. These $\dqcdmu$
will be used to get $\dmkdmu$ in the below.

\begin{figure}[t!]
\vspace{10pt}
\centerline{\epsfig{file=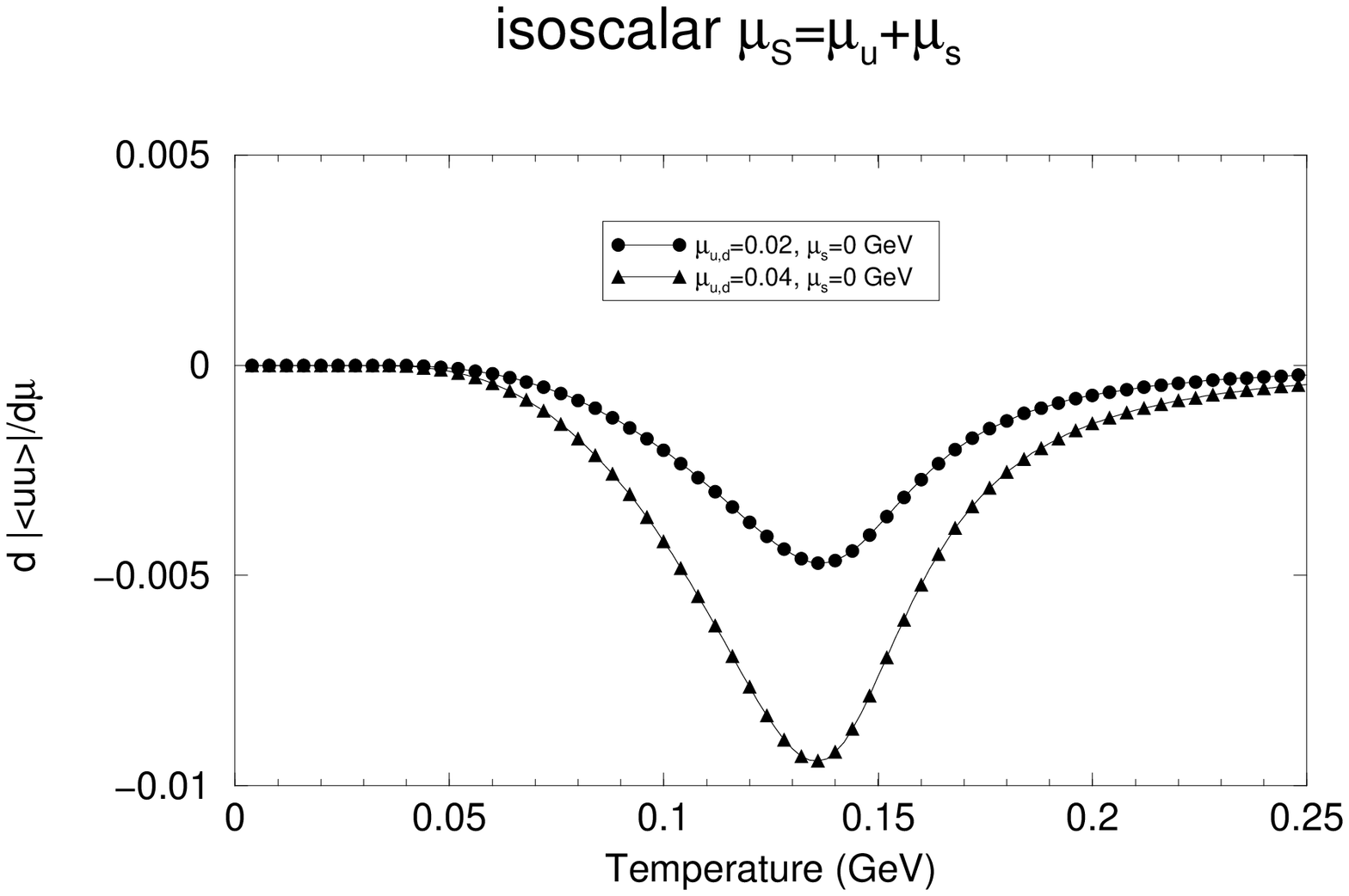,height=7cm,width=7.0cm}
\hspace{1.0cm}\epsfig{file=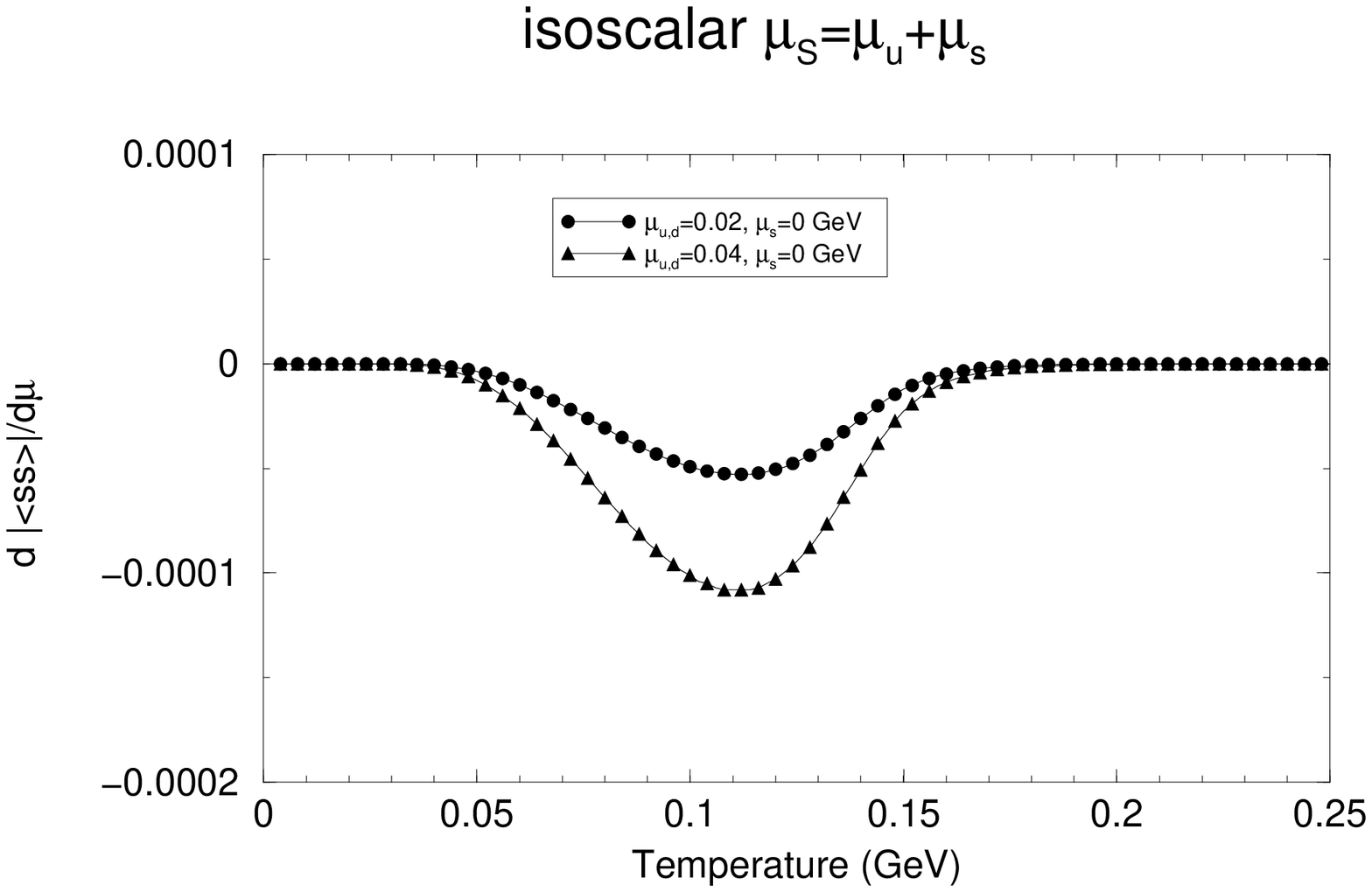,height=7cm,width=7.0cm}}
\vspace{10pt} \caption{The responses of the $u$ (left) and $s$
(right) quark condensates.} \label{dqc}
\end{figure}

Fig.\ref{dqc} shows $\ducdmus$ and $\dscdmus$ at finite chemical
potential. At zero chemical potential both $\dqcdmus$ and
$\dqcdmuv$ are zero. We take two different values for the chemical
potential, $\mu_u$ = $\mu_d$ = 0.02 and 0.04 GeV. In the figure we
set the perpendicular axis as the absolute value of $\dqcdmus$ ,
i.e. ${\partial |\qc| \over \partial \mu_S}$, and thus the figure
shows that the absolute value of the quark condensate decreases
with increasing chemical potential. In addition, the figure shows
that variations of the $u$ quark condensate $\ducdmus$ are much
larger than those of $\dscdmus$, and the variation of each quark
condensate is proportional to the chemical potential.

\begin{figure}[b!]
\vspace{10pt}
\centerline{\epsfig{file=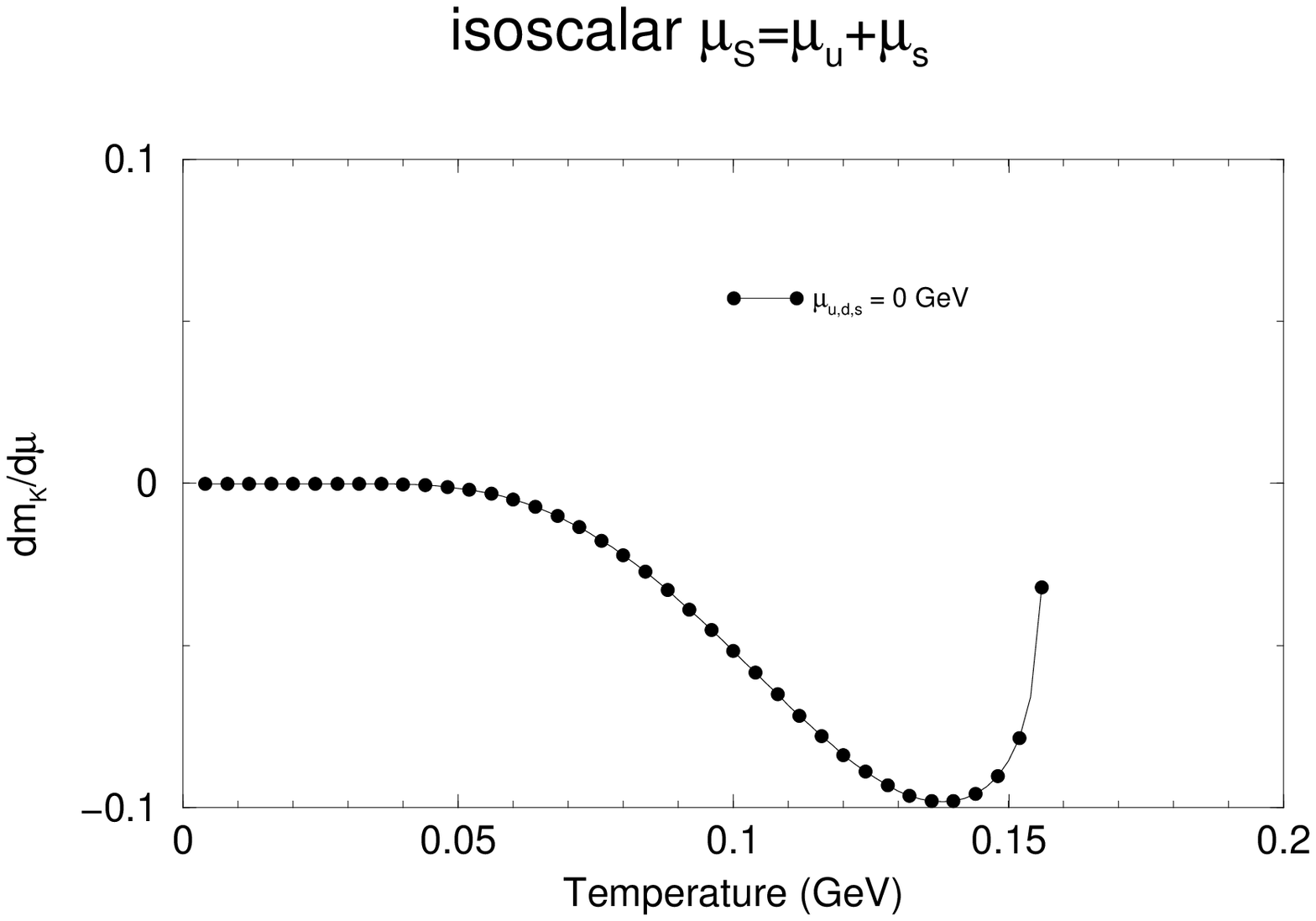,height=7cm,width=7.0cm}}
\vspace{10pt} \caption{$\dmkdmus$ for the $K^-$ at zero chemical
potential.} \label{isos-k-zero}
\end{figure}
\begin{figure}[t!]
\vspace{10pt}
\centerline{\epsfig{file=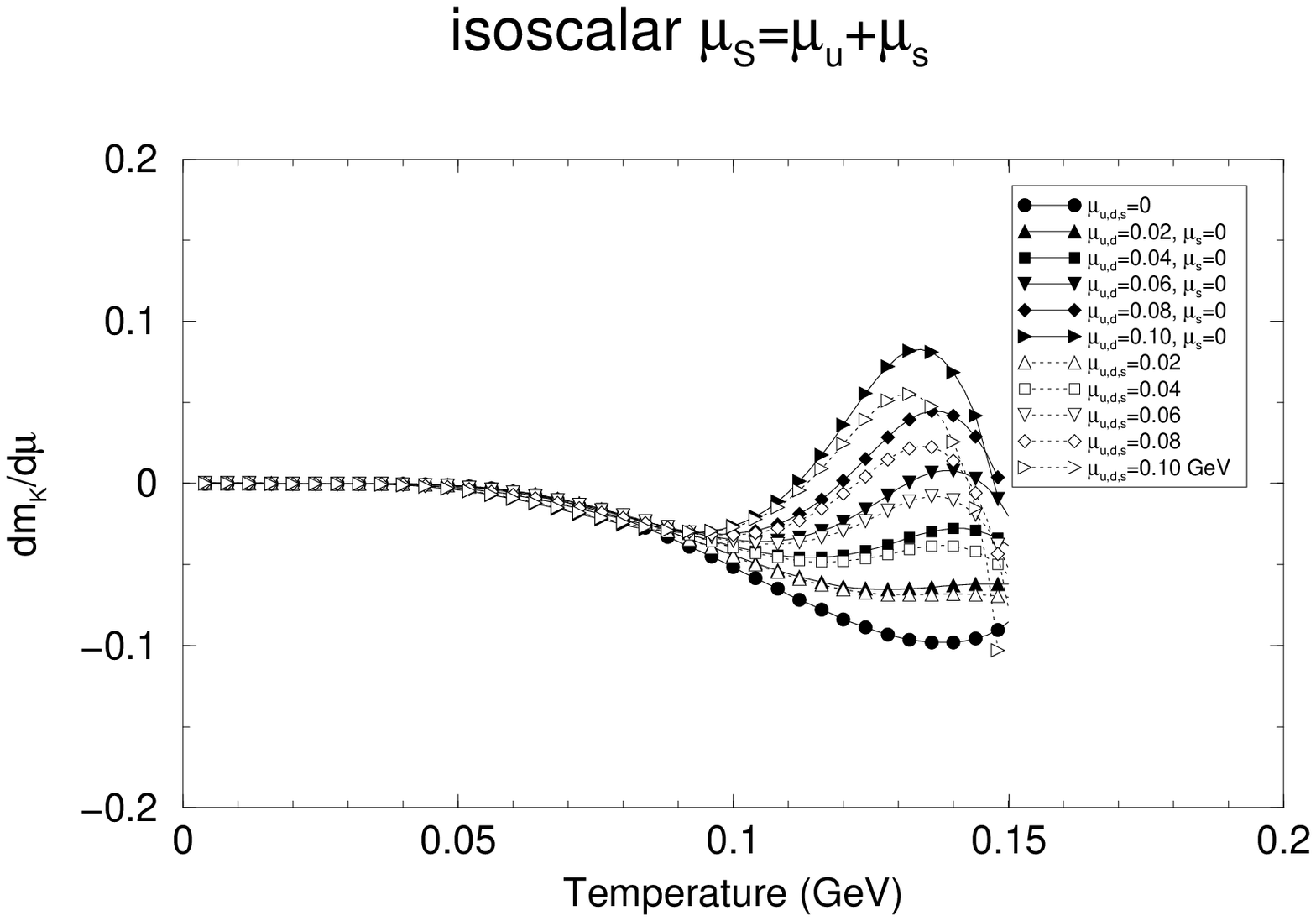,height=7cm,width=7.0cm}
\hspace{1.0cm}\epsfig{file=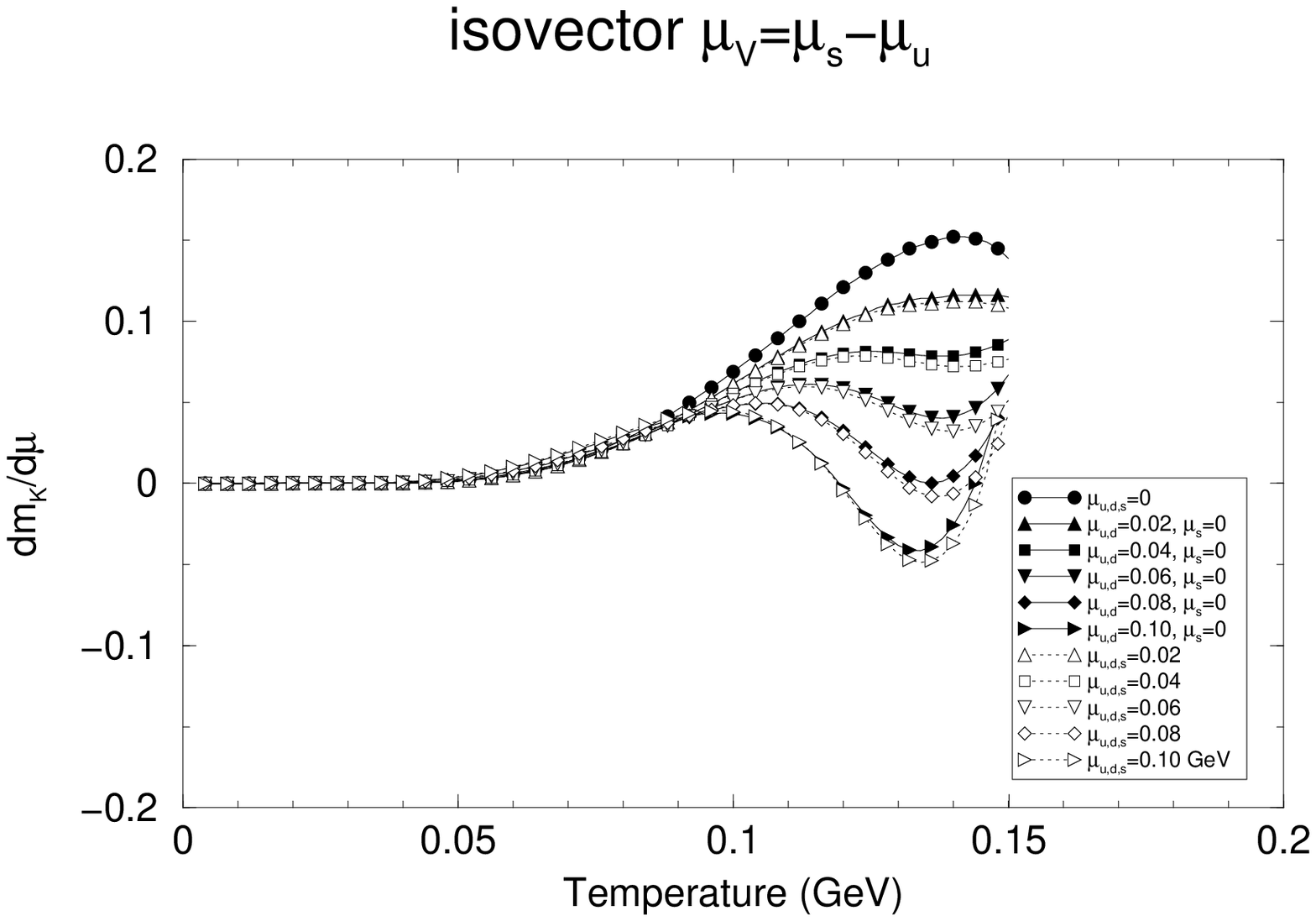,height=7cm,width=7.0cm}}
\vspace{10pt} \caption{$\dmkdmu$ for the $K^-$.} \label{isosv-k}
\end{figure}
\begin{figure}[b!]
\vspace{10pt}
\centerline{\epsfig{file=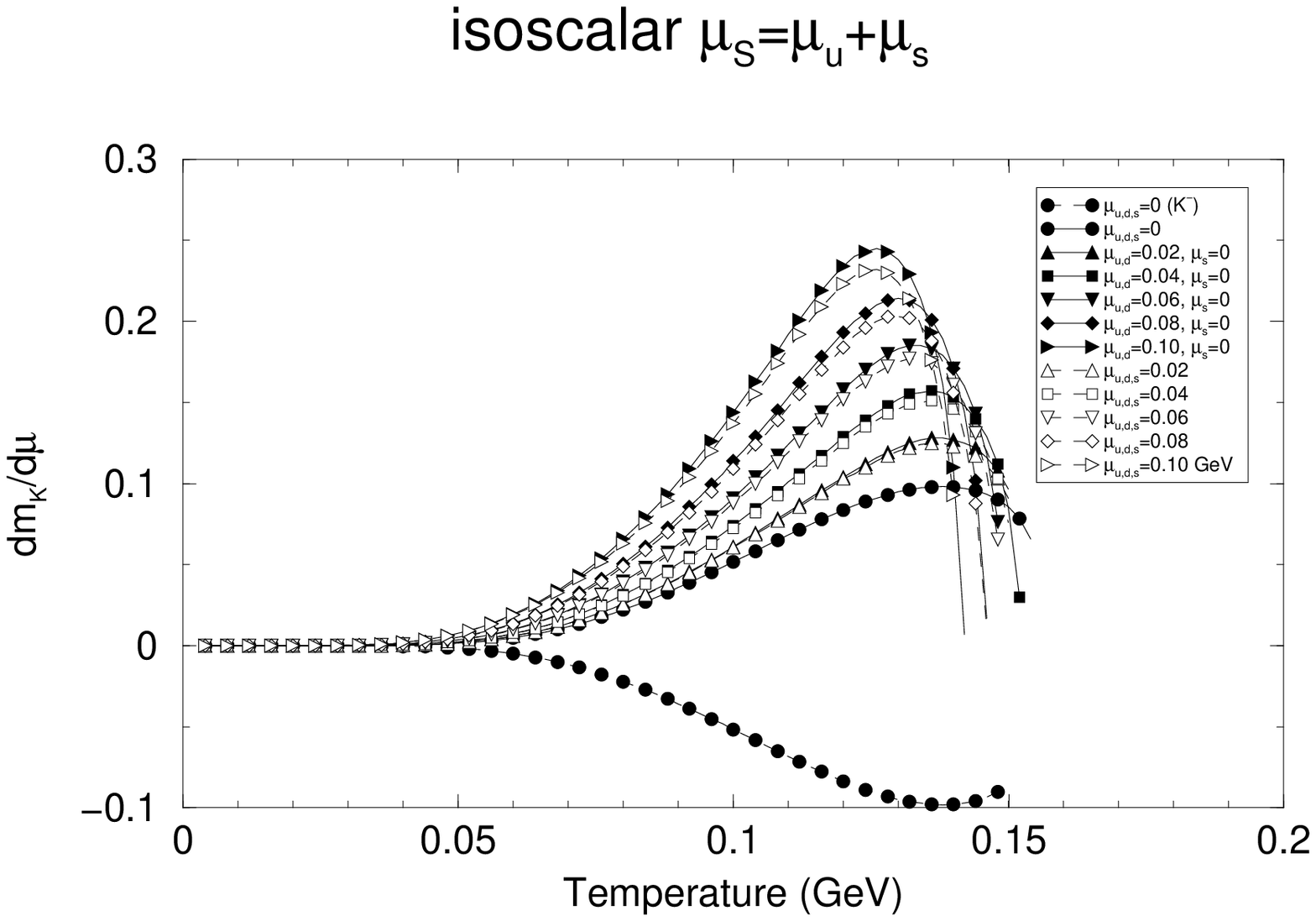,height=7cm,width=7.0cm}
\hspace{1.0cm}\epsfig{file=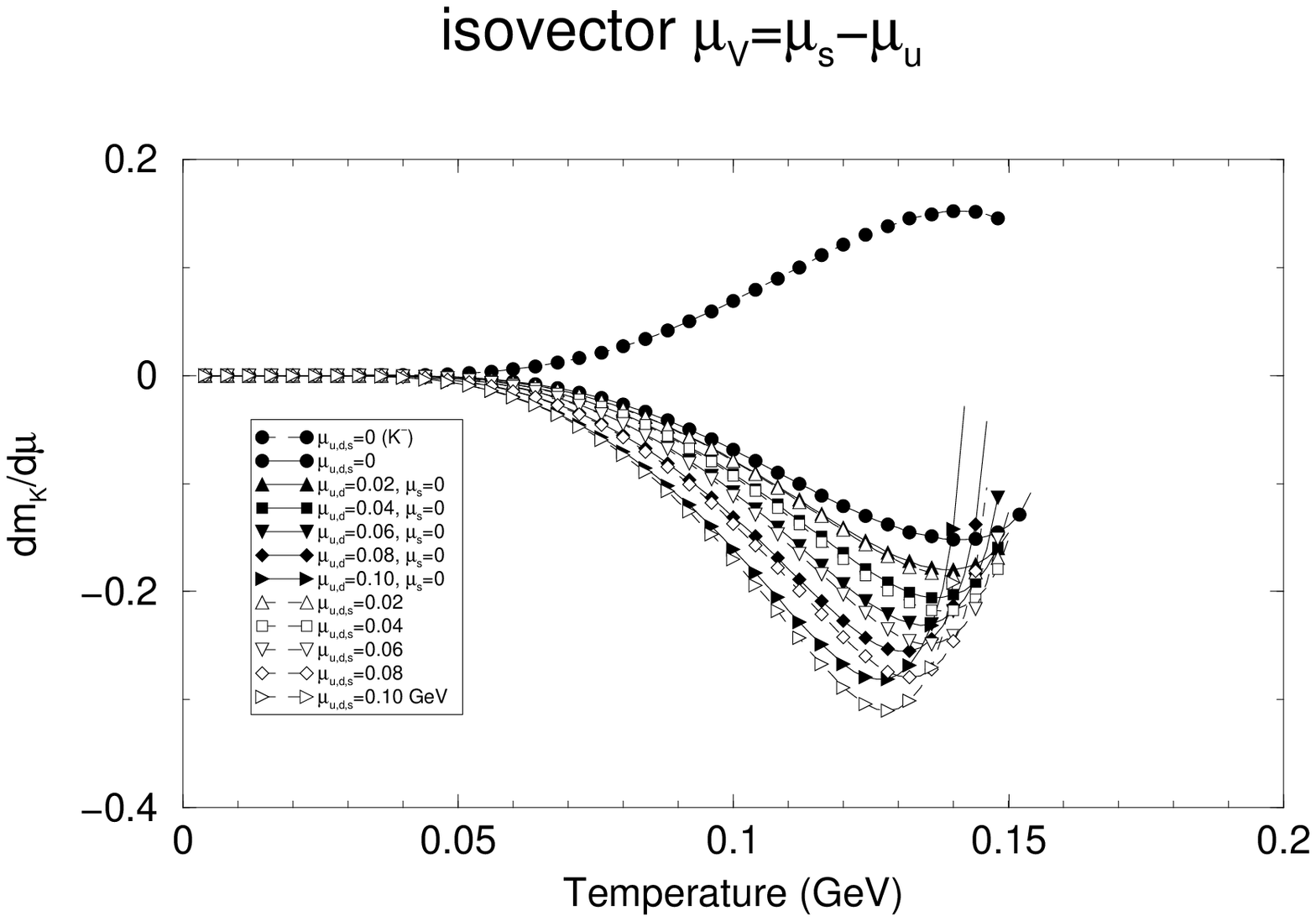,height=7cm,width=7.0cm}}
\vspace{10pt} \caption{$\dmkdmu$ for the $K^+$.}
\label{isosv-kplus}
\end{figure}

Now, consider the dispersion equation for the kaon, e.g., the
$K^-$ \cite{hk94}.
\beq
 D^R_{K^-}(\omega, \vec{q}=0)^{-1} \equiv - G_K^{-1} \left[ 1
+ 2 G_K I^p_{su}(\omega, \vec{q}=0) \right] = 0  \ ,
\label{dispersion}
 \eeq
where $G_K$ is the coupling strength in this channel, $G_K \equiv
g_S + g_D \beta$, and $I^p_{su}(\omega, \vec{q}=0)$ is the
one-loop polarization due to $u$- and $s$-quarks. Differentiating
both sides of the above equation with respect to $\mu_S$ (or
$\mu_V$) at the fixed $T$, and using $\dqcdmus$ (or $\dqcdmuv$) we
get $\dmkdmus$ (or $\dmkdmuv$), i.e. the response of the kaon mass
to changes in the isoscalar (or isovector) chemical potential
$\mu_S$ (or $\mu_V$).

First, we show $\dmkdmus$ for the $K^-$ at zero chemical potential
in Fig. \ref{isos-k-zero}. Below $T$ $\sim$ 0.04 GeV $\dmkdmus$ is
almost zero, and this is because $\dqcdmus$ is hardly changed in
this region as shown in Fig. \ref{dqc}. Near the kaon Mott
temperature $T_{m_K}$ $\dmkdmus$ changes rapidly and becomes
almost zero. Here, $T_{m_K}$ is defined as a temperature at which
the sum of the $u$ and $s$ constituent quark masses equals to the
kaon mass, i.e. $M_u + M_s = m_K$. Above $T_{m_K}$ the kaon
becomes a resonance.

In the figure we do not show the points in the above $T_{m_K}$
region because there may be a large uncertainty. We can not get a
reliable kaon mass in this region, and hence $\dmkdmus$. In fact,
the authors of \cite{hk94} presented the kaon mass in this region
using the imaginary part of the self-energy. However, the
imaginary part is an artifact of the model and thus we need
physical justifications before using this part. In this work we
take only the real part and concentrate on the below $T_{m_K}$
region.

Fig.\ref{isosv-k} shows $\dmkdmus$ and $\dmkdmuv$ at zero and
finite chemical potential. It shows that $\dmkdmus$ increases with
increasing chemical potential, and there is a critical value
between $\mu_u$ = $\mu_s$ = 0.06 and 0.08 GeV where the sign of
$\dmkdmus$ is changed even at below $T_{m_K}$. This result is
consistent with previous NJL model calculations
 \cite{rs9697}. As in the case of zero chemical potential
$\dmkdmus$ changes rapidly near $T_{m_K}$. Now, consider the
isovector case, where $\mu_V = \mu_s - \mu_u$. Then, one can
expect that the sign of $\dmkdmuv$ will be opposite to that of
$\dmkdmus$ because the $u$ quark plays a dominant role rather than
the $s$ quark does. $\dmkdmuv$ decreases with increasing chemical
potential as shown in the figure.

In Fig. \ref{isosv-kplus} we present $\dmkdmus$ and $\dmkdmuv$ for
the $K^+$. They are obtained by replacing $\omega$ in Eq.
(\ref{dispersion}) with $- \omega$. For comparison we also show
$\dmkdmus$ and $\dmkdmuv$ for the $K^-$ at zero chemical
potential.

\section*{3. $\dmpidmu$ in the NJL model}
\label{sec3}

In this section we show $\dmpidmu$ for the $\pi^-$ and $\pi^+$. We
use the same formulas in the previous section by replacing $m_s$,
$\mu_s$ with $m_d$, $\mu_d$, respectively. As for the dispersion
equation a new coupling strength $G_\pi \equiv g_S + g_D \gamma$
is introduced \cite{hk94}.

\begin{figure}[b!]
\vspace{10pt}
\centerline{\epsfig{file=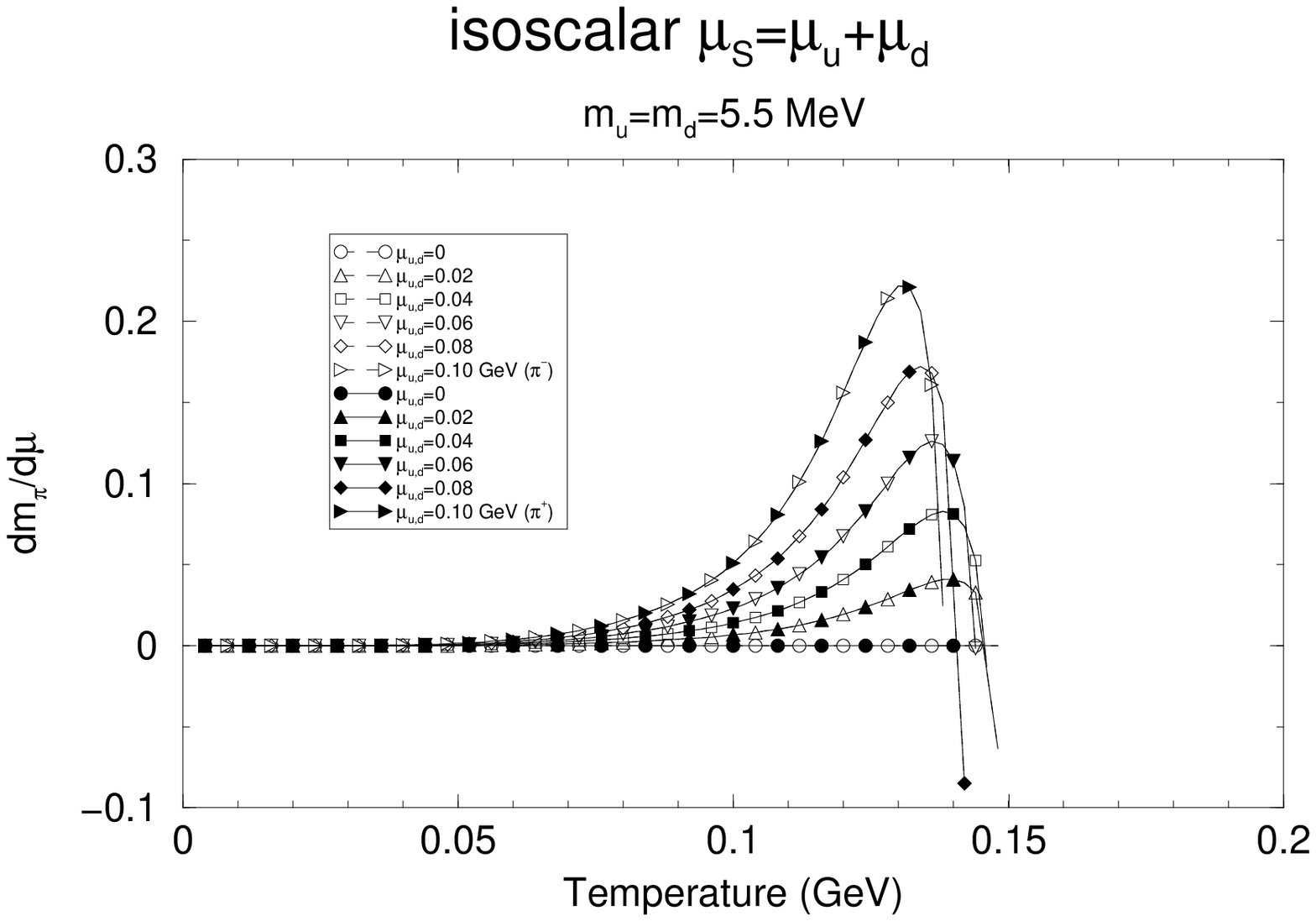,height=7cm,width=7.0cm}
\hspace{1.0cm}\epsfig{file=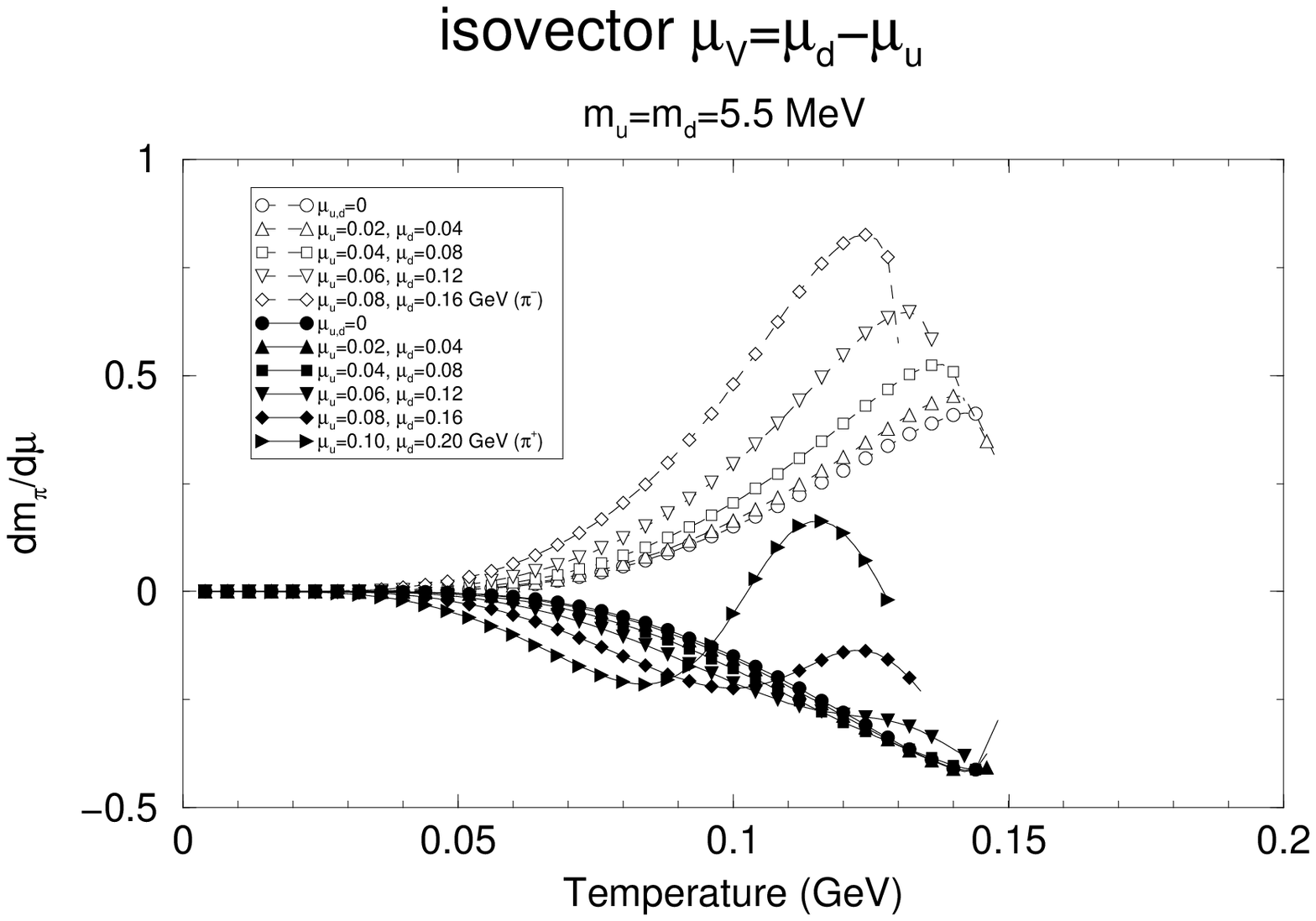,height=7cm,width=7.0cm}}
\vspace{10pt} \caption{$\dmpidmu$ for the pion with $m_u$ = $m_d$
= 5.5 MeV.} \label{isosv-p1plus}
\end{figure}

First, consider the isoscalar $\mu_S = \mu_u + \mu_d$. In the case
$m_u$ = $m_d$, $\dmpidmus$ = 0 at zero chemical potential.
$\dmpidmus$ and $\dmpidmuv$ for the $\pi^-$ and $\pi^+$ at finite
chemical potential are given in Fig.\ref{isosv-p1plus}. Note that
in the case of the isovector chemical potential we take $\mu_d$ =
2 $\mu_u$.

In the previous calculation we assumed $m_u$ = $m_d$ = 5.5 MeV. It
will be interesting to consider different $u$ and $d$ quark
masses, e.g., $m_u$ = 4 MeV and $m_d$ = 7 MeV. Although the
cut-off and the coupling constants should be modified according to
this change of the quark masses, we use the same parameters as
before and study $\dmpidmus$ and $\dmpidmuv$.

\begin{figure}[t!]
\vspace{10pt}
\centerline{\epsfig{file=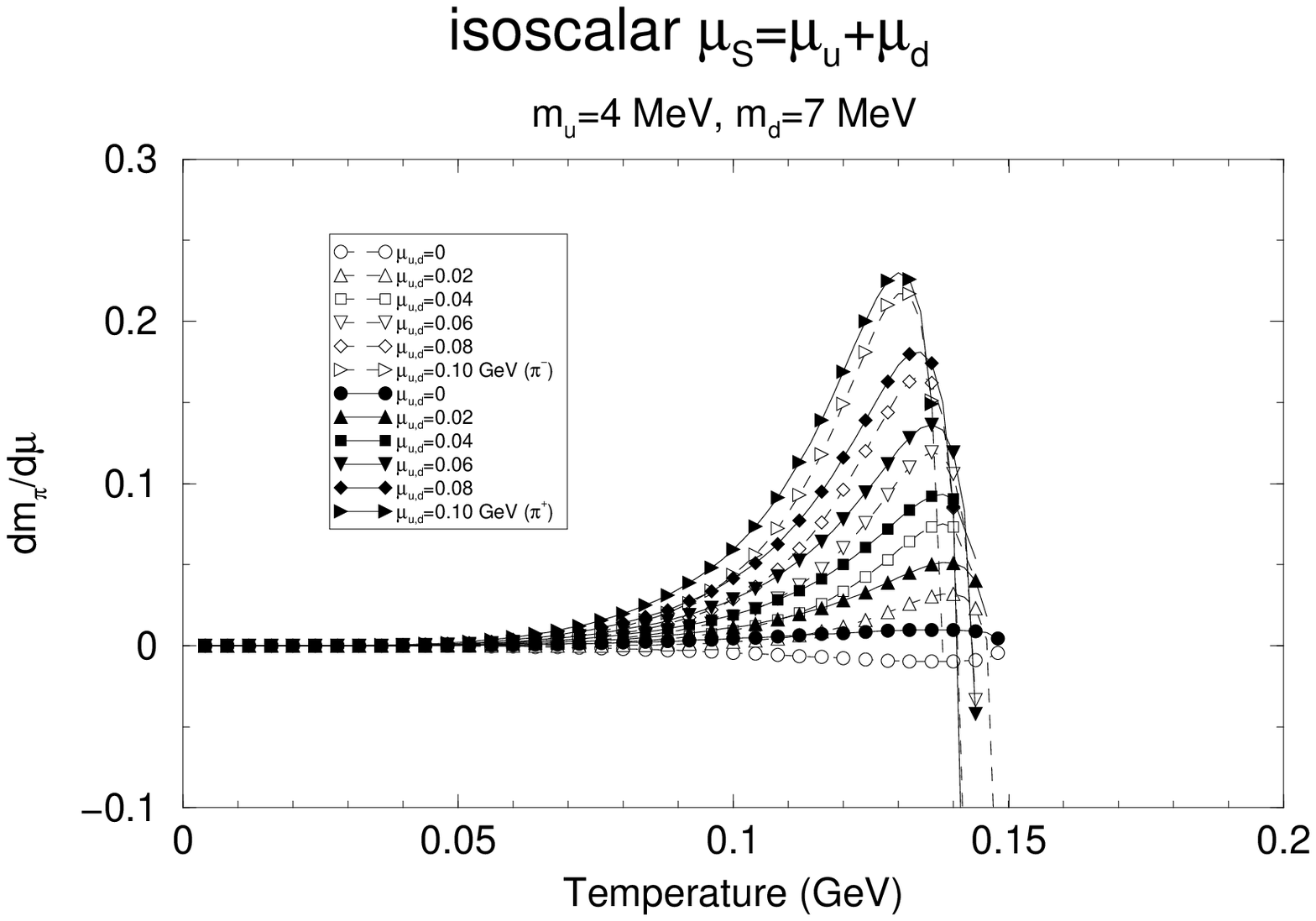,height=7cm,width=7.0cm}
\hspace{1.0cm}\epsfig{file=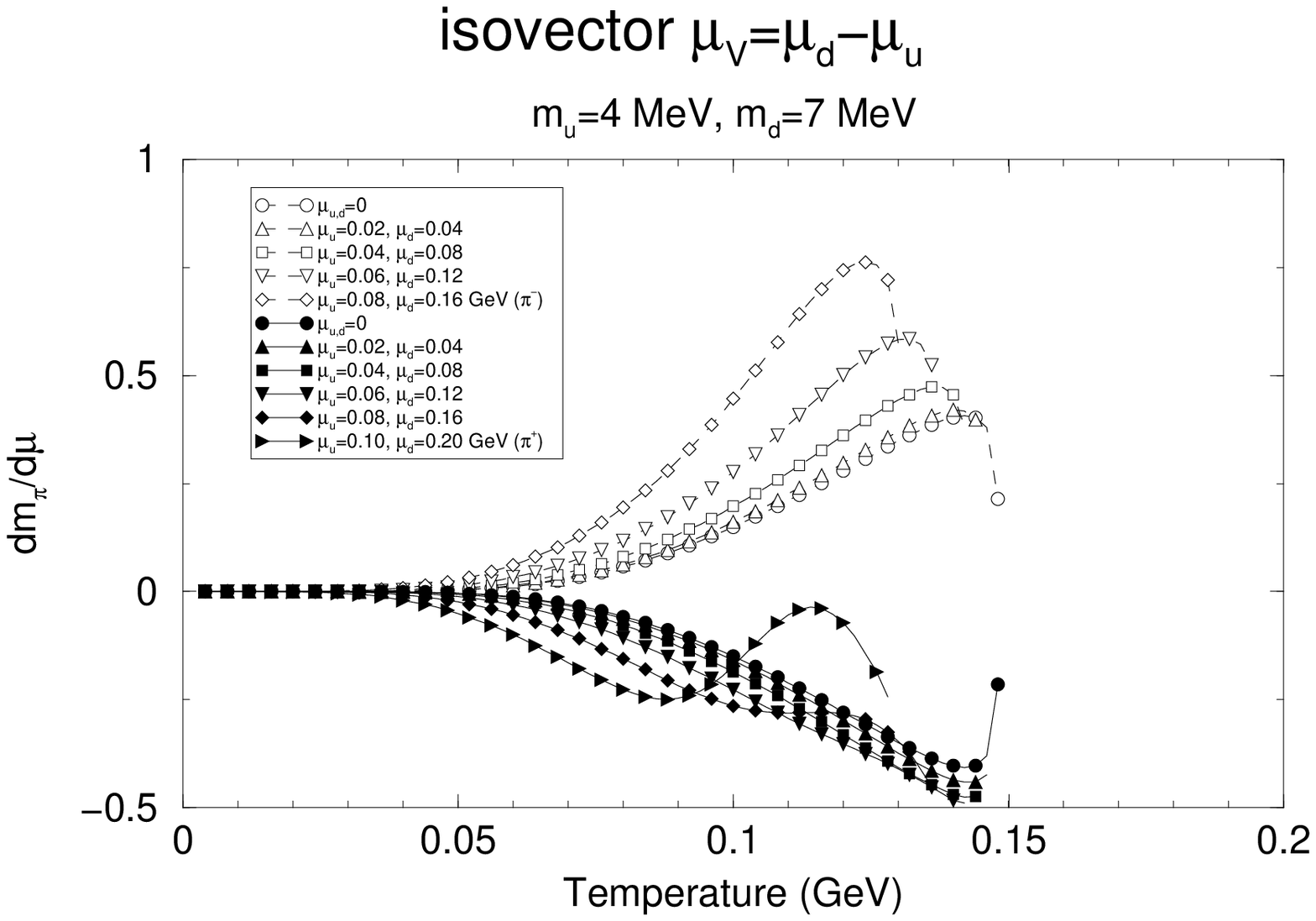,height=7cm,width=7.0cm}}
\vspace{10pt} \caption{$\dmpidmu$ for the pion with $m_u$ = 4 MeV,
$m_d$ = 7 MeV.} \label{isosv-p2plus}
\end{figure}

In the case of $\dmpidmus$ for the $\pi^-$ a transition point
appears between $\mu_u$ = $\mu_d$ = 0.004 GeV and 0.006 GeV as
shown in Fig. \ref{isosv-p2plus}. This transition point seems
reasonable considering the mass ratios of $m_s / m_u$ for the kaon
and $m_d / m_u$ for the pion. For comparison we show both
$\dmpidmus$ for the $\pi^-$ and the $\pi^+$. On the other hand, in
the case of the isovector chemical potential $\dmpidmuv$ for the
$\pi^-$ and $\pi^+$ are similar to the previous ones, i.e. the
results for the pion with the degenerate $u$ and $d$ quark masses.

\section*{4. Discussions}
\label{sec4}

Using the NJL model we have calculated responses of the kaon and
pion masses to changes in the chemical potential, $\dmkdmu$ and
$\dmpidmu$, at zero and finite chemical potential, and found that
$\dmdmu$ is much dependent on the mass difference of two quarks,
i.e. the mass difference between the $u$ and $s$ (or $d$) quarks.

Let us discuss some uncertainties in our calculations. First, we
have considered the Lagrangian (Eq.(\ref{lag})) without the vector
and axial-vector terms. Although there are still arguments about
the strength of the vector coupling $g_V$ \cite{gv}, a further
analysis including these terms is required. In fact, one of the
NJL model calculations showed that the $K^-$ mass at finite
density with $g_V \neq$ 0 is quite different from that with $g_V$
= 0 \cite{rsp99}. A preliminary result of $\dmkdmus$ for the $K^-$
with a non-zero $g_V$ also confirms this \cite{mc01}.

Second, in the previous section we have also considered the
different $u$, $d$ quark masses for the pion ($m_u$ = 4 MeV and
$m_d$ = 7 MeV) and assumed the other parameters are invariant
under this change, and found that in the case of $\dmpidmus$ the
result is slightly different from the previous one, i.e. the pion
with the degenerate $u$ and $d$ quark masses ($m_u$ = $m_d$ = 5.5
MeV). However, we have to take into account variations of the
cut-off and coupling constants, although we expect that they would
be very small. In the real world, SU(2) symmetry is slightly
broken ($m_u \neq m_d$, $\uc \neq \dc$), thus a more careful
analysis is needed in this case.

Third, in this work we have mainly considered the Case II in
\cite{hk94}, where only $g_D$ has the temperature dependence as
shown in Eq.(\ref{gd}). It may be interesting to compare $\dmkdmu$
and $\dmpidmu$ for the Case II with those for the Case I, where
all the coupling constants ($g_S$, $g_D$) and the cut-off
$\Lambda$ are independent of temperature and/or chemical
potential. We have checked that the behaviors of $\dmkdmu$ and
$\dmpidmu$ for the Case I are similar to those for the Case II
except for the different Mott temperatures \cite{mc01}. This is
because $g_D$ is rather irrelevant to the pion and kaon masses.
However, further analyses including all variations of the cut-off
and coupling constants at finite temperature and/or chemical
potential are required before any firm conclusions may be drawn.

As a final remark, we find that the second order responses of the
kaon and pion masses to the chemical potential, $\ddmkdmu$ and
$\ddmpidmu$, are much larger than $\dmkdmu$ and $\dmpidmu$,
respectively. Thus, one can see rather clearer signals than
before.

\section*{Acknowledgements}

We thank T. Kunihiro, K. Redlich, T. Hatsuda, and Su H. Lee for
valuable comments. The work of O.M. was supported by Grant-in-Aide
for Scientific Research by Monbusho, Japan (No. 11694085 and No.
11740159), and the work of S.C. was supported by the Japan Society
for the Promotion of Science (JSPS).

%

%

%
\end{document}